\documentclass[twocolumn,showpacs,preprintnumbers,amsmath,amssymb]{revtex4}

\newcommand{\be}{\begin{equation}}
\newcommand{\ee}{\end{equation}}
\newcommand{\bea}{\begin{eqnarray}}
\newcommand{\eea}{\end{eqnarray}}

\newcommand{\nn}{\nonumber}
\newcommand{\nl}{\nonumber \\}
\usepackage{graphicx}
\usepackage{dcolumn}
\usepackage{bm}
\input epsf.sty

\begin{document}

\preprint{APS/123-QED}

\title{Equilibria with incompressible flows from symmetry analysis\footnote{Published in Phys. Plasmas {\bf 22}, 084502 (2015).}
\\
}


\author{Ap Kuiroukidis$^{1}$ and G. N. Throumoulopoulos$^{2}$}
 \altaffiliation[E-mail: kouirouki@astro.auth.gr,$\; \; $gthroum@cc.uoi.gr ]{}
\affiliation{%
$^{1}$Technological Education Institute of Serres, 62124 Serres, Greece
\\
$^{2}$ University of Ioannina, Department of  Physics,
 GR 451 10 Ioannina, Greece
}%


\date{\today}


\begin{abstract}

We identify and study new nonlinear axisymmetric  equilibria
with incompressible flow of arbitrary direction satisfying a generalized Grad Shafranov equation by extending the symmetry analysis  presented in [G. Cicogna and F. Pegoraro, Phys. Plasmas {\bf 22}, 022520 (2015)]. In particular, we construct a typical tokamak D-shaped equilibrium with  peaked toroidal current density, monotonically varying safety factor and sheared electric field.

\end{abstract}

\maketitle



The analysis of symmetry properties of ordinary or partial
differential equations is a very useful and fruitful tool for studying
the general structure and the space of solutions and for finding
explicit solutions. We consider symmetries described by continuous
Lie groups of transformations, see for example \cite{olve}-\cite{blum}. In particular such symmetry techniques were applied to construct linear and nonlinear solutions of the Grad-Shafranov equation \cite{ka}-\cite{ci}.

This paper is concerned with the study of a class of solutions
obtained through Lie-group-symmetry analysis of the following generalized
Grad-Shafranov  equation (GGS)
\cite{tass}-\cite{thtapo}:
\bea (1-M_p^2) \Delta^\star \psi -
         \frac{1}{2}(M_p^2)^\prime |\nabla \psi|^2
                     + \frac{1}{2}\left(\frac{X^2}{1-M_p^2}\right)^\prime & & \nl
+\mu_0 r^2 P_s^\prime + \mu_0 \frac{r^4}{2}\left[ \frac{\rho(\Phi^\prime)^2}{1-M_p^2}\right]^\prime
    = 0 &&
                            \label{GGS}
 \eea
Here,
   the
  poloidal magnetic flux function $\psi(r,z)$  labels the magnetic surfaces,
  where  ($r,\phi, z$) are cylindrical coordinates with $z$ corresponding to the axis of symmetry;
 $M_p(\psi)$ is
 the   Mach function of the poloidal fluid velocity with respect to the
poloidal  Alfv\'en velocity;
 $X(\psi)$ relates to the toroidal magnetic
 field, $B_\phi=I/R$,  through $I=X/(1-M^{2})$; $\Phi(\psi)$ is the electrostatic potential;
 for vanishing flow the surface function $P_s(\psi)$
  coincides with the pressure; $B$ is the magnetic field modulus
  which can be expressed in terms of surface functions and $R$;  $\Delta^\star=R^2\nabla\cdot(\nabla/R^2)$;
  and the prime denotes derivatives  with respect to $\psi$.
  Because of incompressibility the density $\rho(\psi)$ is also a surface quantity and the Bernoulli equation for the pressure decouples from (\ref{GGS}):
\be
 P=P_s(\psi) -  \rho\left[\frac{v^2}{2}-\frac{r^2(\Phi^\prime)^2}{1-M_p^2} \right]
                          \label{pres}
 \ee
where $v$ is the velocity modulus. The  quantities
$M_p(\psi)$,  $X(\psi)$,
$P_s(\psi)$, $\rho(\psi)$ and $\Phi(\psi)$ are free functions. Derivation of  (\ref{GGS}) and (\ref{pres}) is
 provided  in  \cite{tass}-\cite{thtapo}.

Eq. (\ref{GGS}) can be  simplified  by  the  transformation
\begin{equation}
u(\psi) = \int_{0}^{\psi}\left\lbrack 1 -
M_p^{2}(f)\right\rbrack^{1/2} df
                                            \label{trans}
\end{equation}
under which  (\ref{GGS})  becomes
\bea
  \Delta^\star u
+ \frac{1}{2}\frac{d}{du}\left(\frac{X^2}{1-M_p^2}\right)  +
\mu_0 r^2\frac{d P_s}{d u} & & \nl
+ \mu_0\frac{r^4}{2}\frac{d}{du}\left[\rho\left(\frac{d \Phi}{du}\right)^2\right] = 0 & &
                            \label{GGSt}
\eea
 Note  that no quadratic term as $|{\bf\nabla}u|^{2}$ appears
anymore in (\ref{GGSt}). It is emphasized that once a solution of (\ref{GGSt}) is obtained,  the equilibrium can be completely  constructed with calculations in the $u$-space by employing (\ref{trans}), and  the inverse transformation
\begin{equation}
\psi(u) = \int_{0 }^{u}\left\lbrack 1 -
M_p^{2}(f)\right\rbrack^{-1/2} df
                                            \label{intrans}
\end{equation}
For example, one has for the electric field
$$
{\bf E}=-\nabla \Phi= -\frac{d \Phi(\psi)}{d \psi} \nabla \psi= -\frac{d \Phi(u)}{d u} \nabla u
$$
Contrary to what is stated in \cite{cico2} the (explicit) inversion $\psi=\psi(u)$ is not needed provided that $M_p^2$ and the other free surface quantities  are assigned as function of $u$.
For parallel flows ($\Phi^\prime=0$),  Eq.  (\ref{GGSt}) reduces in form to  the usual GS equation.

The symmetry properties of (\ref{GGSt}), expressed in terms of alternative surface quantities (Eqs. (13)-(16) of \cite{cico2}) in connection with the variational  derivation of \cite{anmo}, were studied in \cite{cico2} (see also  \cite{cico1} and \cite{gebh}).
Henceforth, we follow closely the analysis developed in  \cite{cico2} and extend the
results therein.


Choosing  the free function terms in (\ref{GGSt}) as
\bea
\frac{1}{2}\frac{d}{du}\left(\frac{X^2}{1-M_p^2}\right)=\frac{a_0}{u^3}, \   \ \mu_0 \frac{d P_s}{d u}=\frac{a_2}{u^7},  & & \nl
 \mu_0\frac{d}{du}\left[\rho\left(\frac{d \Phi}{du}\right)^2\right]=\frac{a_4}{u^{11}} & &
                                    \label{choice}
\eea
where $a_0,a_2,a_4$ are free parameters,
 (\ref{GGSt}) assumes the form
\be
\Delta^\star u = \frac{a_0}{u^3}+ r^2 \frac{a_2}{u^7} + r^4 \frac{a_4}{u^{11}}
                               \label{ggs2}
\ee
This equation admits the Lie point scaling symmetry
\bea
\label{symm1}
X_{1}=r\frac{\partial}{\partial r}+z\frac{\partial}{\partial z}
+\frac{1}{2}\psi\frac{\partial}{\partial\psi}
\eea
and the `exceptional' symmetry
\bea
\label{symm2}
X_{2}=2rz\frac{\partial}{\partial r}+(z^{2}-r^{2})\frac{\partial}{\partial z}
+z\psi\frac{\partial}{\partial\psi}
\eea
If $Y=z\frac{\partial}{\partial r}-\sigma r\frac{\partial}{\partial z}$ is
introduced as a weaker type of conditional symmetry (see \cite{cico1,cico2}) then we can write the GGS equation in terms of the $Y-$invariant variable
$s=\sigma r^{2}+z^{2},\; \; (\sigma \neq 0, 1)$. Then   the above mentioned
symmetries map solutions into solutions of the form
\bea
\label{solut}
u(r,z)=s_{\lambda}^{1/4}\phi(y_{\lambda};\alpha_{0},\alpha_{2},\alpha_{4})
\\
y_{\lambda}:=\frac{r^{2}}{s_{\lambda}},\; \; \;
s_{\lambda}:=\sigma r^{2}+\left[z+\lambda (r^{2}+z^{2})\right]^{2} \nn
\eea
For $\sigma=-1$,  (\ref{solut}) recovers  the class of solutions obtained in \cite{cico2} without the restriction of constant densities adopted therein;   solutions of the form (\ref{solut}) hold for arbitrary Mach functions, $M_p^2(u)$ and densities $\rho(u)$.

We will construct and study one tokamak pertinent solution of this symmetry-generated class. Substituting
 (\ref{solut}) into (\ref{ggs2}) we obtain
\bea
\label{ode}
[4\sigma(\sigma-1)y^{3}+4(1-2\sigma)y^{2}+4y]\phi^{''}+\nn \\
+[6\sigma(\sigma-1)y^{2}+(4-6\sigma)y]\phi^{'}+\nn \\
+[-\frac{1}{4}+\frac{3\sigma(1-\sigma)}{4}y]\phi=\frac{\alpha_{0}}{\phi^{3}}
+\frac{\alpha_{2}y}{\phi^{7}}+\frac{\alpha_{4}y^{2}}{\phi^{11}}
\eea



In order that  the coefficient of $\phi^{''}$ in (\ref{ode})
vanishes for a particular $y=y_0$, we choose $y_{0}=1/(\sigma-1)$.  Then, to integrate  (\ref{ode}) we employ the initial conditions
$\phi(y_{0})=1$ and
\bea
\label{deriv}
4y_{0}\phi^{'}(y_{0})=[\frac{1+3\sigma}{4}+\alpha_{0}+\alpha_{2}y_{0}+\alpha_{4}y_{0}^{2}]
\eea
the latter one stemming from (\ref{ode}) for $y=y_0$.
 We solve numerically  (\ref{ode}) by  choosing
$\sigma=1/4$, $\lambda=\sqrt{3}/4$, $\alpha_{0}=1/4$, $\alpha_{2}=-0.5$,
 $\alpha_{4}=0.1$ and integrating in the interval $y_{0}\leq y\leq {\bar y}_{0}$,
with ${\bar y}_{0}=1/\sigma$. Using cubic fitting we found that
\bea
\label{fit}
\phi(y)=-1.341y^{3}-4.343y^{2}-4.999y-1.106
\eea
Thus, we have obtained the equilibrium configuration  with a crescent-shaped
cross-section shown in  Fig. 1. By exploiting the invariance of GGS under constant displacements $z\rightarrow z+d$, the configuration has been  properly displaced along the z-axis   so that the magnetic axis lies on the plane $z=0$.
The outermost magnetic surface corresponding to the flux
value of $u_{out}=-125$ Wb  touches the $z$-axis  at a couple of corners.
At the magnetic axis   located at $r_{0}=1.19$, $z_{0}=0.05$  we have $u_{0}=-135$ Wb.
A more pertinent tokamak  equilibrium can be constructed by choosing a
fixed boundary to coincide with an interior D-shaped magnetic surface. Such a  boundary corresponding to the
value of $u_{b}=-132.543$ Wb is indicated by the black-colored curve in Fig. 1 and separately in Fig. 2.

\begin{figure}[ht!]
\centerline{\mbox {\epsfxsize=10.cm \epsfysize=8.cm \epsfbox{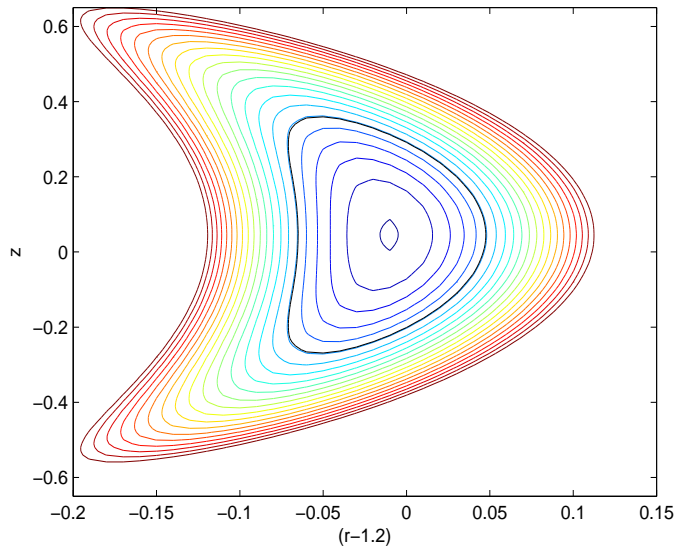}}}
\caption[]{$u$-contours on the poloidal plane for an equilibrium of the class of solutions (\ref{solut}).
The black-colored D-shaped curve corresponds to the boundary of a tokamak pertinent equilibrium shown separately in Fig. 2.}
\label{fig1}
\end{figure}

\begin{figure}[ht!]
\centerline{\mbox {\epsfxsize=10.cm \epsfysize=8.cm \epsfbox{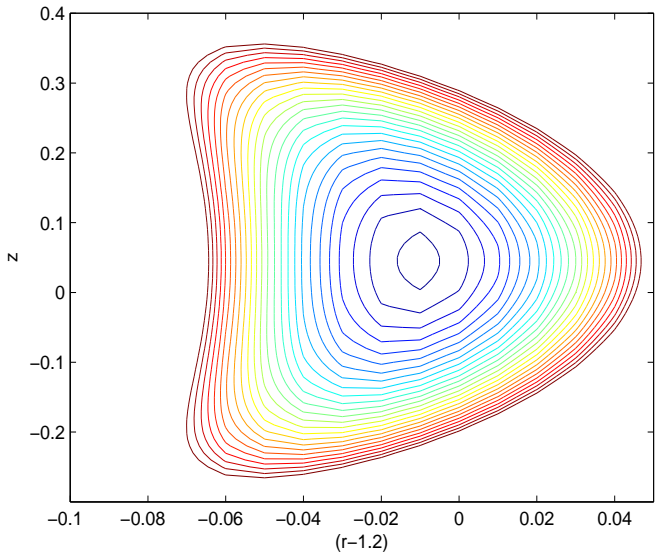}}}
\caption[]{$u$-contours on the poloidal plane for the tokamak   D-shaped equilibrium.}
\label{fig2}
\end{figure}

\begin{figure}[ht!]
\centerline{\mbox {\epsfxsize=10.cm \epsfysize=8.cm \epsfbox{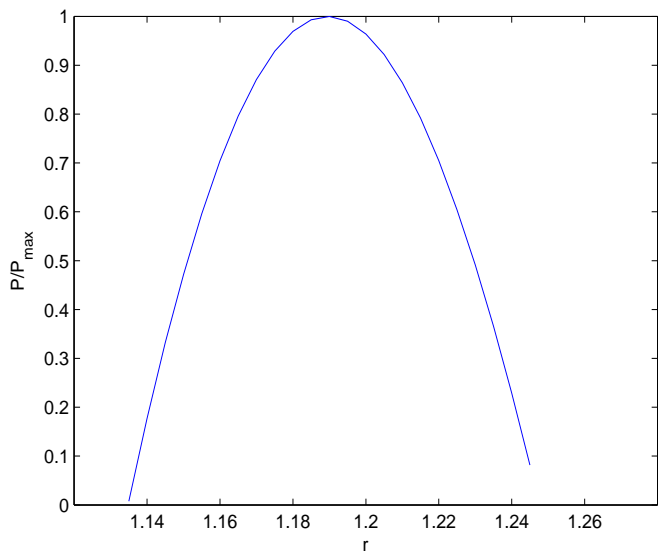}}}
\caption[]{The pressure profile on the plane $z=0$ for the D-shaped equilibrium  of   Fig. 2.
It vanishes on the boundary of the D-shaped equilibrium at $u_{b}=-132.543$ Wb. Here
$P_{max}=0.316$ Atm.}
\label{fig3}
\end{figure}

\begin{figure}[ht!]
\centerline{\mbox {\epsfxsize=10.cm \epsfysize=8.cm \epsfbox{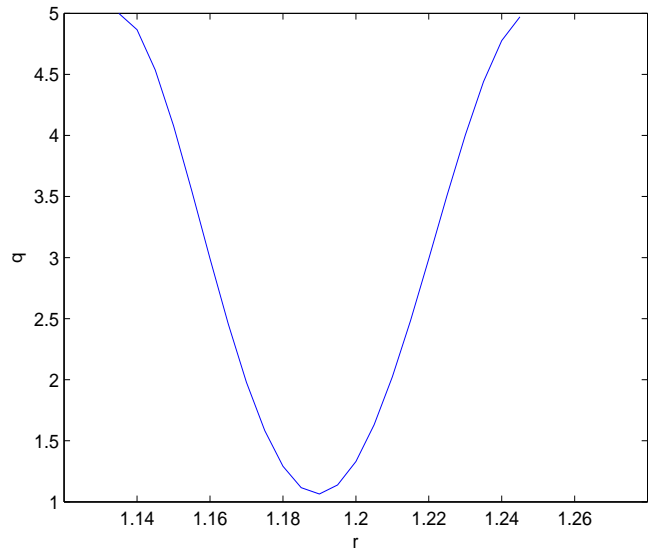}}}
\caption[]{The safety factor for the D-shaped equilibrium of Fig. 2.}
\label{fig4}
\end{figure}

\begin{figure}[ht!]
\centerline{\mbox {\epsfxsize=10.cm \epsfysize=8.cm \epsfbox{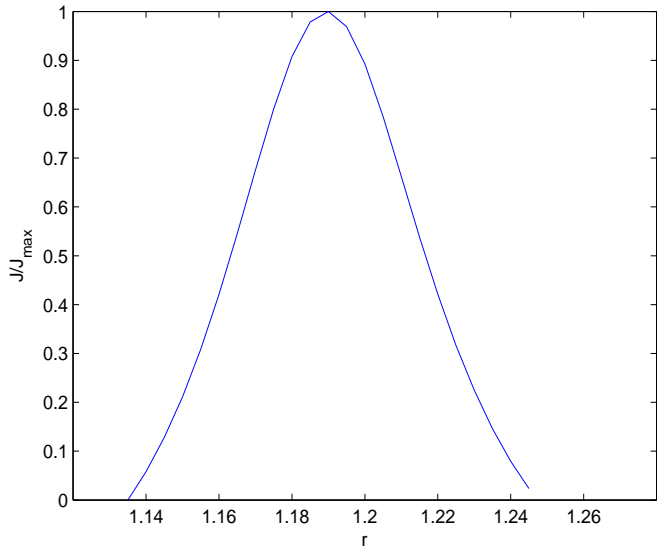}}}
\caption[]{The current density   for the D-shaped equilibrium of Fig. 2. It vanishes on
the boundary of the D-shaped equilibrium. Here $J_{max}\simeq 1.7 MA/m^{2}$.}
\label{fig5}
\end{figure}

\begin{figure}[ht!]
\centerline{\mbox {\epsfxsize=10.cm \epsfysize=8.cm \epsfbox{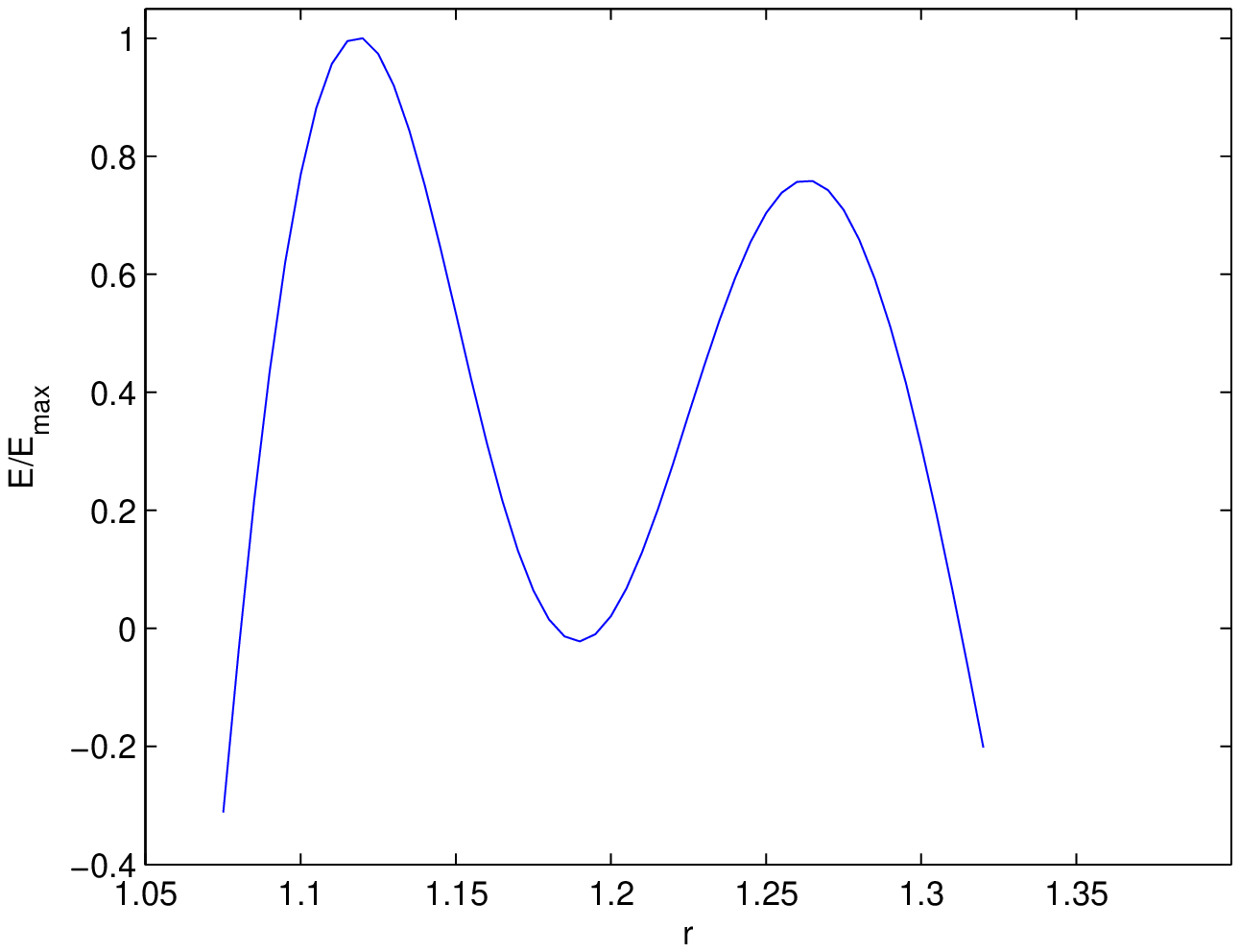}}}
\caption[]{The electric field for the   D-shaped equilibrium of Fig. 2. It vanishes on axis. Here $E_{max}=$ 1 Kv/m.}
\label{fig6}
\end{figure}

To completely  construct the equilibrium we choose the Mach-function and density peaked on the magnetic axis and vanishing on the boundary as
$M_{p}^{2}(u)=M_{0}^{2}(1-u^{2}/u_{b}^{2})$, $\rho=\rho_{0}(1-u^{2}/u_{b}^{2})$ respectively,
where $M_0^=0.02$ and $\rho_0=4\times 10^{-7}$ Kg/m$^{3}$.
The  pressure of the D-shaped  equilibrium of Fig. 2, peaked on axis and vanishes on the
boundary,  is shown in Fig. 3. The safety factor  has a typical tokamak variation
monotonically increasing from the magnetic axis to the boundary, as can be  seen
in Fig. 4. The toroidal current density is also peaked on axis and vanishing on the boundary (Fig. 5) unlike a
hollow  $J_\phi$ profile  of an equilibrium with parallel flow corresponding to
the same ansatz (\ref{choice}) with $a_4=0$ constructed in \cite{kuth}. This result
indicates that the electric field associated with the non-parallel component
of the flow, shown in Fig. 6, may play a role in determining the equilibrium characteristics.

In summary, by employing  Lie and weak conditional symmetries  of a  GGS  equation  for plasmas with incompressible flow of arbitrary direction we have extended   a class of nonlinear axisymmetric   solutions
obtained
in  \cite{cico2}. In particular, we  constructed a D-shaped configuration with sheared electric field and typical tokamak equilibrium characteristics, i.e.   pressure and toroidal  current density peaked on the magnetic axis and safety factor monotonically increasing from the axis to the boundary.  It would be interesting to  extend the search for
other relevant equilibria by potentially identifying other
weak  conditional symmetries  following the procedure of   {\cite{cico1} and {\cite{cico2}.
Work along these lines is in progress.



\section*{Aknowledgments}\

This work has been carried out within the framework of
the EUROfusion Consortium and has received funding from
the National Programme for the Controlled Thermonuclear
Fusion, Hellenic Republic. The views and opinions expressed
herein do not necessarily reflect those of the European
Commission.


\newpage

\bibliography{apssamp}

\end{document}